\documentclass[preprint]{elsarticle}
\usepackage{xspace}
\usepackage{graphicx}
\usepackage{amssymb,amsmath}
\usepackage{subfig}
\usepackage{color}
\newcommand{\nc}[1]{\newcommand #1}
\newcommand{\rnc}[1]{\renewcommand #1}
\nc{\myquote}[1]{`#1'}
\nc{\x}[1]{\mbox{#1}}
\rnc{\matrix}[2]{\left[\!\!\begin{array}{#1} #2\end{array}\!\!\right]}
\rnc{\vector}[1]{\matrix{c}{#1}}
\nc{\suml}[2]{\sum \limits_{#1}^{#2}}
\nc{\real}[1]{\Re\lbrace #1 \rbrace}
\nc{\imag}[1]{\Im\lbrace #1 \rbrace}
\nc{\g}[1]{\x{$#1$}}
\nc{\e}[2]{\begin{equation} #1 \label {eq:#2} \end{equation}}
\nc{\ea}[2]{
\begin{eqnarray}
#1 \label {eq:#2} \end{eqnarray}}
\nc{\inv}{^{-1}}
\nc{\tra}{^{\mathrm T}}
\nc{\herm}{^{\mathrm H}}
\nc{\fabstand}{\,}
\nc{\fp}{\fabstand.}
\nc{\fk}{\fabstand,}
\nc{\mm}[1]{\mathbf{#1}}
\nc{\mms}[1]{\boldsymbol{#1}}
\nc{\ie}{i.\,e.\xspace}
\nc{\eg}{e.\,g.\xspace}
\nc{\cf}{cf.\xspace}
\nc{\ndim}{N_{\mathrm{dim}}}
\nc{\nh}{H}
\nc{\ntd}{N_{{t}}}
\nc{\niter}{N_{\mathrm{iter}}}
\nc{\nfeval}{N_{\mathrm{eval}}}
\nc{\hbm}{HB}
\nc{\hbmwocond}{\hbm}
\nc{\dd}{{\mathrm{d}}}
\nc{\ii}{{\mathrm{i}}}
\nc{\ee}{{\mathrm{e}}}
\nc{\deltamp}{\xi}
\nc{\basefreq}{\Omega}
\nc{\basefreqst}{\Omega^*}
\nc{\umax}{u_{\mathrm{max}}}
\nc{\qmod}{{q}}
\nc{\amod}{{a}}
\nc{\amodst}{{a^*}}
\nc{\ommod}{{\omega_0}}
\nc{\ommodst}{{\omega_0^*}}
\nc{\dmod}{D}
\nc{\dmodtil}{\tilde{D}}
\nc{\omex}{{\Omega_{\mathrm e}}}
\nc{\omexst}{{\Omega_{\mathrm e}^*}}
\nc{\omone}{{\omega_1}}
\nc{\omtwo}{{\omega_2}}
\nc{\omonesq}{{\omega_1^2}}
\nc{\omtwosq}{{\omega_2^2}}
\nc{\done}{{\zeta_1}}
\nc{\dtwo}{{\zeta_2}}
\nc{\uonest}{{u_1^*}}
\nc{\uonedotst}{{\dot u_1^*}}
\nc{\utwost}{{u_2^*}}
\nc{\utwodotst}{{\dot u_2^*}}
\nc{\umodtwost}{{u_{{m}=2}^*}}
\nc{\umodtwodotst}{{\dot u_{{m}=2}^*}}
\nc{\umodsixst}{u_{m=6}^*}
\nc{\umodsixdotst}{{\dot u_{m=6}^*}}
\nc{\utwoonehat}{\hat u_{m=2,n=1}}
\nc{\absphase}{\theta}
\nc{\phaselag}{\Delta\theta}
\nc{\phaselagd}{\Delta{\dot\theta}}
\nc{\phiex}{\phi_{\mathrm e}}
\nc{\phiexd}{\dot\phi_{\mathrm e}}
\nc{\femod}{\hat f_{\mathrm{e,mod}}}
\nc{\ql}{\hat q}
\nc{\phil}{\mm\phi}
\nc{\oml}{\omega}
\nc{\qlk}{\ql_k}
\nc{\philk}{\phil_k}
\nc{\omlk}{\oml_k}
\nc{\cnm}{\mm Y}
\nc{\uhn}{\mm{\hat u}_n}
\nc{\psihs}{\hat\psi}
\nc{\psih}{\mms{\psihs}}
\nc{\psihn}{\mms{\psihs}_n}
\nc{\uhone}{\mm{\hat u}_1}
\nc{\fnls}{f}
\nc{\fnl}{\mm \fnls}
\nc{\fex}{\mm e}
\nc{\ehone}{\hat{\fex}_1}
\nc{\fnln}{{\fnl}^{\mathrm N}}
\nc{\fnlh}{\hat{\fnl}}
\nc{\kt}{k_{\mathrm{t}}}
\nc{\npp}{N_{\mathrm p}^*}
\nc{\fdder}{\nabla}
\nc{\fr}{f_{\mathrm R}}
\nc{\fref}[1]{\x{Fig.~\ref{fig:#1}}}
\nc{\eref}[1]{\x{Eq.~(\ref{eq:#1})}}
\nc{\erefo}[1]{(\ref{eq:#1})}
\nc{\tref}[1]{\x{Table~\ref{tab:#1}}}
\nc{\sref}[1]{\x{Section~\ref{sec:#1}}}
\nc{\srefo}[1]{\ref{sec:#1}}
\nc{\srefs}[1]{\x{Sections~\ref{sec:#1}}}
\nc{\zo}[1]{\x{\cite{#1}}}
\nc{\fig}[3][tbh]{
\begin{figure}[#1]
\centering
\includegraphics{figs/#2}
\caption{#3\label{fig:#2}}
\end{figure}}
\nc{\figw}[3][tbh]{
\begin{figure}[#1]
\centering
\includegraphics[width=1.0\textwidth]{figs/#2}
\caption{#3\label{fig:#2}}
\end{figure}}
\nc{\tab}[5][tbh]{
\begin{table}[#1]
\centering
\caption{#4\label{tab:#5}}
\begin{tabular}{#2}
\hline
#3 \\
\hline
\end{tabular}
\end{table}}
\begin{document}

\begin{frontmatter}

\title{Nonlinear modal analysis of
nonconservative systems: Extension of the periodic motion
concept}

\author[ids]{Malte Krack\corref{cor1}}
\ead{krack@ila.uni-stuttgart.de}


\cortext[cor1]{Corresponding author}

\begin{abstract}
As the motions of nonconservative autonomous systems are typically not periodic, the definition of nonlinear modes as periodic motions cannot be applied in the classical sense. In this paper, it is proposed 'make the motions periodic' by introducing an additional damping term of appropriate sign and magnitude. It is shown that this generalized definition is particularly suited to reflect the periodic vibration behavior induced by harmonic external forcing or negative linear damping. In a large range, the energy dependence of modal frequency, damping ratio and stability is reproduced well. The limitation to isolated or weakly-damped modes is discussed.
\end{abstract}


\end{frontmatter}

\section{Introduction}
The concept of nonlinear modes is an attempt to extend the ideas of
linear modal analysis to nonlinear systems. Nonlinear modal analysis
is useful to characterize and quantify the energy dependence of the
essential vibration behavior of nonlinear systems. This facilitates
the understanding of phenomena like energy transfer, localization
and modal interactions that may be caused by nonlinear effects.
Moreover, nonlinear modes can be used for the purpose of model
reduction. To this end, the problem is reduced to the typically
two-dimensional subspace spanned by a specific nonlinear mode. This
can greatly improve the computational efficiency of vibration
predictions, which is crucial \eg for design optimization and
uncertainty analysis involving detailed models of nonlinear
structures \zo{krac2014a,krac2014b}. For recent overviews on the
general topic, the interested reader is referred to
\zo{kers2009,vaka2008}.\\
Useful mathematical properties are lost when nonlinear effects
become important. For this reason, the definition of nonlinear
modes\footnote{In literature, the term Nonlinear Normal Mode (NNM)
is quite common. However, the term \myquote{normal} may mislead to
the wrong conclusion that nonlinear modes are orthogonal to each
other. Apparently this term goes back to Rosenberg \zo{rose1960},
who defined nonlinear modes as vibrations in \textit{unison}, \ie
where all material points cross their equilibrium point and their
extremum points simultaneously. For this type of vibration, the
motions take place on so-called modal lines in the generalized
displacement space which are normal to the surface of maximum
potential energy \zo{vaka2008}. However, this property is only valid
for symmetric conservative systems. Hence the term \myquote{normal}
in this context is avoided throughout this article.} is less
straight forward. There are basically two different established
definitions: (a) the periodic motion definition and (b) the
invariant manifold definition.\\
According to definition (a), nonlinear modes are viewed as periodic
motions of the autonomous nonlinear system \zo{rose1960,rand1974}. A
family of periodic motions can be defined that are continuations
with respect to the kinetic energy. These branches of periodic solutions
of the equation of motion may or may not be connected to a
corresponding linear mode at low energies. By using continuation and
bifurcation analysis, interactions between different nonlinear modes
can be resolved. The periodic motion definition directly associates
properties of the flow to the modes such as its frequency and its asymptotic
stability. Unfortunately, it does not apply to nonconservative systems, since their autonomous
motions are typically only periodic at possible limit cycles, \ie at specific energy values. Being inspired by the modes of damped linear systems, Laxalde and Thouverez \zo{laxa2009} defined nonlinear modes as pseudo-periodic motions. They computed them by means of the generalized Fourier-Galerkin method. In contrast to the oscillatory term in the usual Fourier ansatz, they considered an additional
decay term. This approach is, however,
strictly limited to trigonometric base functions, and the notion of
mode stability has not been established yet.\\
According to definition (b), nonlinear modes are viewed as an
invariant relationship between the coordinates of an autonomous
system. This relationship defines a manifold in the system's phase
space that includes the equilibrium point, where it is tangential to the
hyper-plane spanning the locus of the corresponding linear mode
\zo{shaw1991,shaw1993,shaw1994}. In the simplest case without internal resonances, this manifold is two-dimensional.
Hence, a point on the manifold can be parameterized uniquely by a
suitable set of two coordinates. Only the geometry of the invariant
manifold is governed by this definition. To assess the vibratory
properties such as frequency and stability, the in-manifold flow can
be analyzed in a second step. Conceptually, the invariant manifold
definition is not limited to conservative systems. Compared to
periodic motion based methods, two important difficulties are often
reported in practice: the difficulty to find a suitable set of
coordinates for a unique parametrization and the computational
burden. The former difficulty
often arises in conjunction with localization and modal interactions
at higher energies. These phenomena can lead to a folding of the
invariant manifold in certain coordinate systems, so that not every point on it can be
described uniquely anymore, see \eg \zo{blan2013,renson2014}. In
terms of computational effort, it can generally be stated that periodic motion based procedures tend to
be more efficient than invariant manifold based ones. For
periodic motion based methods, only a single periodic orbit needs to
be computed at the same time. In contrast, often the entire manifold, being the locus of
infinitely many orbits, has to be determined simultaneously in the case of invariant manifold
based methods. This
results in a larger problem dimension and can produce considerable computational effort. To simplify the
problem, the manifold can be divided into annular regions of finite
size. But the unsteady character of the flow of nonconservative
systems generally leads to a coupling of these regions, which
requires special attention \zo{renson2014}.\\
The purpose of this article is to extend the periodic motion
definition to dissipative systems. The general concept
is introduced in \sref{extension}. Two computational implementations
of the extended concept are outlined in \sref{computation}. They
are based on Shooting and Harmonic Balance, respectively. In
\sref{numex}, the capabilities and limitations of the approach are assessed for several numerical
examples. 

\section{Extension of the periodic motion concept to dissipative systems\label{sec:extension}}
This section is divided into three subsections: First, the motivation and general idea for the
extension of the periodic motion concept is presented.
Then, the nonlinear modes are defined in such a way that they are capable of
characterizing this dynamic regime in terms of eigenfrequency, modal
damping ratio and vibrational deflection shape. Finally, analogies
to the method of force appropriation are indicated.

\subsection{Periodic vs. damped motion concept}
Suppose the motions of an autonomous nonlinear system are described
by a finite set of generalized coordinates \g{\mm u} and velocities
\g{\mm{\dot u}}, governed by a set of second-order
ordinary differential equations
\e{\mm M\mm{\ddot u}(t) + \fnl\left(\mm u(t),\mm{\dot u}(t)\right) =
\mm 0\fp}{eqm_autonomous}
Herein, \g{\mm M} is the mass matrix and \g{\fnl} are linear or
nonlinear restoring as well as dissipative forces.
\figw[thb]{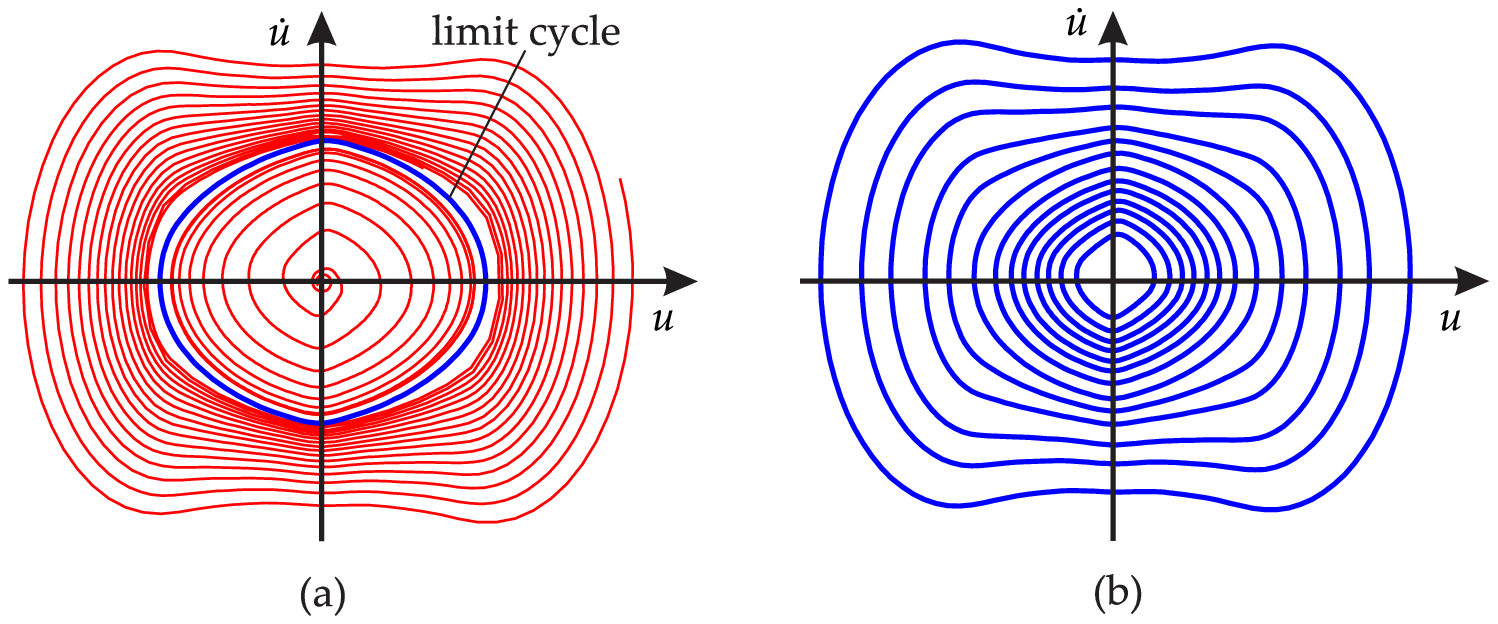}{Visualization of different conceptions
of nonlinear modes of nonconservative systems: (a) conventional
damped motion concept, (b) periodic motion concept followed in
this study}
\\
In the conservative case, the nonlinear modes represent periodic
solutions to \eref{eqm_autonomous}. Once the periodic motion is
initiated, the autonomous system will retain this motion for all
times. This motion can be initiated by according initial conditions.
Also, this motion can be induced by harmonic
external forcing at resonance, when an appropriate term is included into the autonomous \eref{eqm_autonomous}. In other words, the frequency-energy
characteristic of the nonlinear mode is identical to the backbone of
the frequency response curves for varying excitation level. This is
known as the \myquote{deformation-at-resonance} property of
nonlinear modes \zo{szem1979,kers2009}.\\
In the presence of nonconservative forces, an unsteady motion takes
place instead until an attractor is reached, for instance a limit
cycle or an equilibrium point. If the system is subjected to a
harmonic external forcing at resonance, this behavior is changed and
a periodic motion may persist instead. Depending on the excitation level,
different kinetic energy levels can be reached. In the nonconservative case, the
deformation-at-resonance property does not hold anymore, \ie the
backbone curve of the (steady-state) frequency response curves is
not identical to the (instantaneous) frequency-energy characteristic
of the autonomous system. In fact, this deviation is not due to
nonlinearity, but also holds in the linear case: Consider the
behavior of a single-degree-of-freedom oscillator with undamped
eigenfrequency \g{\omega_0} and damping ratio \g{D}. While the
backbone curve is a straight vertical line through the constant
resonance frequency \g{\omega_{\mathrm{res}}=\omega_0\sqrt{1-2D^2}},
the frequency of the autonomous system is the damped
eigenfrequency \g{\omega_{\mathrm d}=\omega_0\sqrt{1-D^2}}.\\
Owing to this peculiarity of nonconservative systems, it is not ad
hoc clear how to define the nonlinear modes. Essentially, there are
two different opportunities:
\begin{itemize}
    \item \textit{damped motion concept}: nonlinear modes shall
    capture the strict autonomous behavior, \ie unsteady motions in
    general, or
    \item \textit{periodic motion concept}:  nonlinear
    modes are still periodic motions, which share, as much as
    possible, the properties of the underlying autonomous system
\end{itemize}
Both conceptions are illustrated in \fref{figure_01}. The
former concept is the classical one and was followed for instance
in \zo{renson2014,laxa2009}. In this study, the latter
concept was followed. More specifically, periodic motions in the presence of (a) harmonic external forcing at
resonance, and (b) self-excitation by negative modal damping represent the dynamic regime
of interest. The aim is therefore to capture the vibration behavior in situations
where a (permanent) \textit{resonant excitation source} is present. The term \myquote{resonant} refers to the fact that in both cases, the oscillation frequency equals one of the system's (energy-dependent) eigenfrequencies. It is believed that this concept is more adjusted to
persistent vibrations of nonconservative systems. Such vibrations
are often of primary concern from a structural dynamic design point of view.

\subsection{A new definition of nonlinear modes}
The remaining question is now how to \textit{make} the motions
periodic. It is proposed to enforce periodicity by an additional damping
term \g{\deltamp\mm M\mm{\dot u}} that is just large enough to
compensate the nonconservative forces,
\e{\mm M\mm{\ddot u}(t) -\deltamp\mm M\mm{\dot u}(t) + \fnl\left(\mm
u(t),\mm{\dot u}(t)\right) = \mm 0\,,\,\, 0\leq t<T\quad\land\quad \mm u(t+T) = \mm u(t)\fp}{eqm_periodic}
The extended definition of nonlinear modes is therefore as follows:
\textit{A nonlinear mode is as a family of periodic motions of an
autonomous nonlinear system. If the system is nonconservative, these
periodic motions are enforced by mass-proportional
damping/self-excitation.}\\
The choice of a mass-proportional damping ensures consistency with
linear modal analysis for the case of symmetric systems with modal damping.
In this case, the nonlinear modes are
orthogonal with respect to the mass matrix. Hence, the damping term
does not affect the mode shape or the natural frequency \g{\ommod}.
The damping term is related to the modal damping by \g{2\dmod\ommod=\deltamp}. In the nonlinear regime,
however, the modes are no longer orthogonal, so that the additional
damping term can induce artificial modal coupling. This makes the
approach intrusive. Owing to this possible source of inaccuracy, the valid
dynamic regime will be restricted to \textit{isolated modes} where
strong modal interactions are absent.
It should also be noted that, by definition, the concept is limited to periodic motions. Consequently the approach is not useful to capture quasi-periodic or even chaotic motions, which may also be encountered near resonance. In general, the autoparametric character induced by the additional term in \eref{eqm_periodic} could trigger the emergence of non-periodic dynamic regimes \zo{Naprstek.2009b,Tondl.2000}.
\\
Akin to their linear counterpart, nonlinear modes are of course a
mathematical concept, regardless whether the conventional or the proposed definition are used. Their
usefulness resides in their ability to reflect the vibration
behavior of nonlinear systems in the dynamic regime of interest. It
is thus crucial to verify this physical significance of the modal
characteristics. This is the objective of the numerical examples in
\sref{numex}.

\subsection{Comparison to force appropriation}
The interested reader might point out that instead of introducing an
additional damping term, one might consider an external harmonic
forcing that drives the system into resonance. An appropriate
condition to ensure resonant behavior could be the phase lag
criterion, which requires a \g{90^{\circ}} phase lag between excitation
and response \zo{peeters2011}. This procedure is known as force
appropriation. Frequency or phase of the (generally multi-harmonic)
external forcing generally have to be adjusted iteratively to meet
the resonance criterion. This adjustment is not required for the method proposed in this study: The correct phase and frequency of the excitation is automatically ensured owing to the positive feedback of the velocity. This is an important advantage of the proposed method.
Moreover, a damping measure \g{\deltamp} or \g{\dmod} is
directly obtained by the modal analysis, whereas such a measure is
not obvious in the case of force appropriation.

\section{Computational procedures\label{sec:computation}}
As the nonlinear modes were defined as periodic motions, standard
methods for analyzing limit cycles can be applied to \eref{eqm_periodic}. Two different computational procedures
are considered for the modal analysis in this study, which are described in the following.
They were based respectively on the well-known Harmonic Balance and the Shooting method.
\subsection{Harmonic Balance}
For this method, the dynamic variables are expanded in a Fourier
series truncated to harmonic order \g{\nh} with base frequency
\g{\ommod}, \eg
\e{\mm u(t) \approx \amod\real{\suml{n=0}{\nh}\psihn\ee^{\ii n\ommod
t}}\fp}{hbm_ansatz}
Herein, \g{\psihn} are the harmonics of the eigenvector and
\g{\amod} is the modal amplitude. The \g{\psihn} may be complex-valued to allow for a phase lag between the coordinates. Due to this particular choice of
base functions, the approximation is automatically periodic. The harmonic order \g{\nh} should be selected as large as
necessary to ensure sufficient accuracy and as small as possible to
avoid spurious computational effort. The substitution of
\eref{hbm_ansatz} into the equations of motion \erefo{eqm_periodic}
and subsequent Fourier-Galerkin projection gives rise to a set of
nonlinear algebraic equations:
\ea{\nonumber\text{solve} & -\left(n^2\ommod^2+\deltamp n
\ommod\right)\mm M\psihn \amod + \\
\nonumber &\quad\quad \fnlh_n
\left(\ommod,\amod\psih_0,\cdots,\amod\psih_{\nh}\right) =
\mm 0\,\,\forall\,n=0,\cdots,\nh\\
\nonumber\text{subject to} & \psih_1\herm\mm M\psih_1 = 1\fk\quad\imag{\psihs_{1,k}}=0\\
\text{with respect to} &
\lbrace\ommod,\deltamp,\psih_0,\cdots,\psih_{\nh}\rbrace\,\,\mathrm{for}\,\,
\amod\in\left[\amod_{\mathrm{min}},\,\amod_{\mathrm{max}}\right]\fp}{complex_evp}
Just as in the general linear case, two normalization conditions are
imposed in order to overcome the otherwise under-determined problem.
A mass normalization of the fundamental harmonic \g{\psih_1} of the
eigenvector was used in this study, which is also commonly applied
in the linear case. Note that in \eref{complex_evp}, \g{{}\herm} denotes the Hermitian transpose and should not be confused with the harmonic order $\nh$. Any eigensolution can be phase-shifted and will still satisfy \eref{complex_evp}.
A phase normalization is therefore proposed, where the
imaginary part \g{\imag{\psihs_{1,k}}} of an arbitrary component of
the fundamental harmonic eigenvector is set to zero. Applying the normalization to the
fundamental harmonic \g{\psih_1} is reasonable since \g{\psihn=\mm 0} for \g{n\neq 1} in the linear case. As an alternative to the mass
normalization, a normalization with respect to the kinetic energy
can be performed, which is also well-suited for the case of internal
resonances \zo{krac2013a}. It should be noticed that only super-harmonics have been
considered in the formulation of \eref{complex_evp} utilized in this study, although the approach
can be extended in a straightforward manner to account for sub-harmonics.

\subsubsection*{Evaluation of the forces}
\fig[tbh]{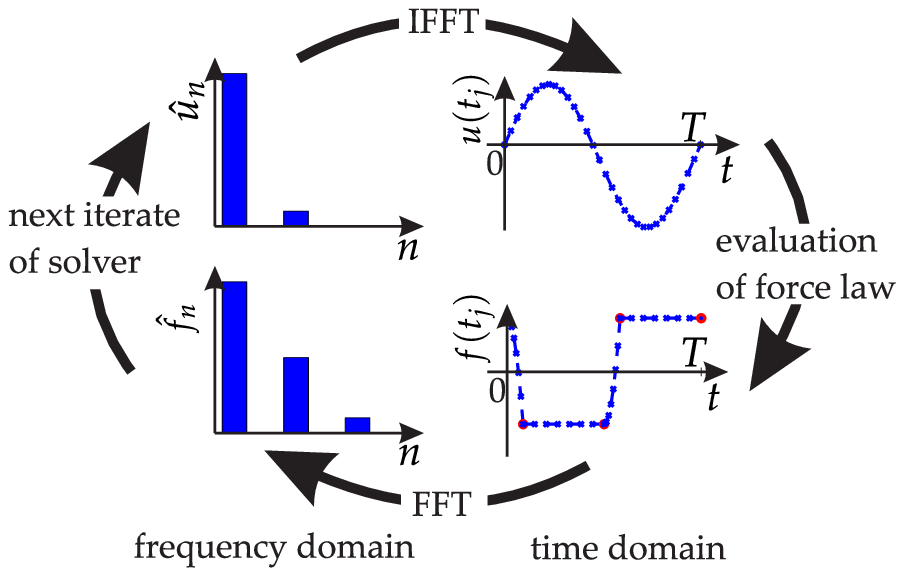}{Alternating-frequency-time scheme}
Since \g{\mm f} is generally nonlinear in \g{\mm u} and \g{\mm{\dot u}}, the harmonic components \g{\fnlh_n} of the nonlinear forces also depend
on the harmonics \g{\amod\psih_1,\cdots,\amod\psih_{\nh}} of the
sought coordinates and on the eigenfrequency \g{\ommod}. In general, this
functional relationship cannot be expressed in closed
form, and numerical procedures have to be utilized for computing \g{\fnlh_n}. The alternating-frequency-time scheme \zo{came1989,cardona1994} is a
widely used technique in this context, and it is illustrated in
\fref{figure_02}. In each iteration of the solver, the \g{\fnlh_n} are evaluated to determine a new estimate for \g{\uhn=\amod\psihn}. The inverse fast fourier transform (IFFT) is applied to determine the generalized coordinates
\g{\mm u(t_j)} and the velocities \g{\mm{\dot u}(t_j)} at equally spaced time
instants \g{t_j}. The force law is then directly evaluated in the time domain
to obtain the force \g{\fnl(t_j)}. Finally, the harmonics \g{\fnlh_n} are obtained by fast fourier transform (FFT). Mathematically, this procedure can be summarized as
\e{\fnlh = \mathrm{FFT}\left[~\mm f\left(\mathrm{IFFT}\left[\mm{\hat u}\right],\,\mathrm{IFFT}\left[\ommod\fdder\mm{\hat u}\right]\right)~\right]\fk}{aft}
with $\mm u=\mathrm{IFFT}\left[\mm{\hat u}\right]$ and $\mm{\dot u}=\mathrm{IFFT}\left[\ommod\fdder\mm{\hat u}\right]$, where $\fdder$ takes care of the differentiation in the frequency domain. This scheme makes it easy to take into account arbitrary nonlinear and even non-smooth forces.

\subsubsection*{Numerical solution}
The eigenproblem in \eref{complex_evp} comprises a number
of \g{\ndim=N\left(2\nh+1\right)+2} unknowns where \g{N} is the number of generalized coordinates. For the numerical
solution, a variant of the Newton method was used. Specifically, the
subroutine \myquote{fsolve} was utilized, which is available in the Matlab software environment. This method requires the computation of the Jacobian, \ie the matrix assembling the derivative of the residual vector with respect to
the vector of unknowns. Throughout this study, the Jacobian was
evaluated using analytical expressions. Typically the critical part is the derivation of the forces $\fnlh$. In the framework of the Alternating-frequency-time scheme, this can be achieved by utilizing the linearity of the FFT and the IFFT operator,
\e{\frac{\partial\fnlh}{\partial x} = \mathrm{FFT}\left[~\frac{\partial\mm f\left(\mm u,\mm{\dot u}\right)}{\partial \mm u}\cdot\mathrm{IFFT}\left[\frac{\partial\mm{\hat u}}{\partial x}\right] +
\frac{\partial\mm f\left(\mm u,\mm{\dot u}\right)}{\partial \mm{\dot u}}\cdot\mathrm{IFFT}\left[\frac{\partial\ommod\fdder\mm{\hat u}}{\partial x}\right]
~\right]\fp}{aft_grad}
Herein, the scalar variable $x$ is one of the unknowns; \ie, it is either $\ommod$, $\deltamp$, or the real or the imaginary part of a harmonic component of a generalized coordinate. The derivation of $\mm f$ with respect to $\mm u$ or $\mm{\dot u}$ is usually straight-forward. The interested reader is referred to \zo{bert1998,siew2009a} more details and examples. After each evaluation of the residual function, thus, also
its Jacobian is available, which is then provided to the nonlinear
solver. This can greatly reduce the required computational cost
compared to the common finite difference approximations of the
Jacobian.
If the system features only localized nonlinearities, the nonlinear
force vector is typically sparse. By exploiting this sparsity, the
computational burden can be further reduced as demonstrated \eg in
\zo{krac2013a}.\\
For the iterative solution procedure, the considered linear mode was
specified as initial guess at small amplitudes \g{\amod\approx 0}.
Turning points with respect to the amplitude may be encountered, for instance in
the presence of internal resonances. In order to overcome these
turning points, a numerical path continuation was employed. The
well-known predictor-corrector scheme with tangent predictor step
and arc-length parametrization was used. The details of this scheme
can be found \eg in \zo{seyd1994}. It should be noticed that more
sophisticated bifurcation analysis techniques could be applied in a
straight-forward manner, this was however considered beyond the
scope of this study.\\
The different unknowns associated with \eref{complex_evp} represent
quite different physical quantities and may assume numerical values
of different order of magnitude. This was found to have a
crucial influence on the convergence behavior of the numerical
solution procedure. A logarithmic scaling was used for the amplitude, as it
typically undergoes comparatively large variations during the
modal analysis. This is accomplished by treating $z:=\log_{10} \amod$ rather than $\amod$ as an unknown, and internally recovering $\amod=10^{z}$, as needed. A linear scaling was applied to
the remaining unknowns, so that they had approximately matching
orders of magnitude.

\subsubsection*{Stability analysis}
Owing to the periodic motion definition, the classical notion of
asymptotic stability of limit cycles can be applied to the nonlinear
modes. For the frequency domain, the Hill method can be employed
\zo{groll2001b,laza2010}. This basically involves the solution of a
quadratic eigenvalue problem. The coefficient
matrices of this problem only depend on quantities readily known from the solution of \eref{complex_evp}. Hence, the stability is typically analyzed a posteriori. Stability is determined by the real part of the Hill eigenvalues: If any of the eigenvalues has a positive real part, the nonlinear mode is unstable at this point.

\subsection{Shooting}
For this method, periodic solutions of \eref{eqm_periodic} are determined by solving the following boundary value problem:
\ea{\nonumber\text{solve} & \vector{\mm u(T)\\ \mm{v}(T)} -
\vector{\mm u_0\\ \mm v_0}=\mm 0\\
\nonumber &\quad \text{where}\, \mm u(T),\,\mm
v(T)\,\text{are obtained from:}\\
\nonumber &\quad\vector{\mm{\dot u}(t)\\\mm{\dot v}(t)} =
\vector{\mm v(t)\\-\mm M\inv\fnl\left(\mm u(t),\mm
v(t)\right)+\deltamp\mm v(t)}\\
\nonumber &\quad \text{with}\,\mm u(0)=\mm
u_0,\,\mm v(0)=\mm v_0\\
\nonumber\text{subject to} & u_{0,k} = \qmod\fk\quad v_{0,k}=0\\
\text{with respect to} &
\lbrace T,\deltamp,\mm u_0, \mm v_0\rbrace\,\,\mathrm{for}\,\,
\qmod\in\left[\qmod_{\mathrm{min}},\,\qmod_{\mathrm{max}}\right]}{shooting_evp}
Herein, \g{\mm v} is the velocity \g{\mm v:=\mm{\dot u}}.
In addition to the initial values \g{\mm
u_0} and \g{\mm v_0}, the period \g{T=\frac{2\pi}{\ommod}} and the damping
coefficient \g{\deltamp} are unknown. For normalization,
displacement \g{u_{0,k}} and velocity \g{v_{0,k}} of a particular coordinate at $t=0$ are
prescribed.

\subsubsection*{Numerical solution}
\eref{shooting_evp} represents a set of nonlinear algebraic
equations, which was solved like \eref{complex_evp} using a variant
of the Newton method in conjunction with a path
continuation procedure. In each iteration of the solver, the states
\g{\mm u(T)} and \g{\mm v(T)} at the end of the period \g{T} are
determined by numerical time-step
integration of \eref{eqm_periodic} 
from the initial states \g{\mm u_0, \mm v_0}. The implicit
average acceleration Newmark scheme was utilized in this study. It was also noted that other methods like the
Runge-Kutta scheme work fine as well. If certain symmetries exist in the temporal evolution, the computational burden of the numerical integration can be reduced, as it is suggested \eg in \zo{peet2009}.\\
It should be remarked that the modal amplitudes \g{\amod} and \g{\qmod} involved
in the Harmonic Balance formulation \erefo{complex_evp} and
the Shooting formulation \erefo{shooting_evp}, respectively,
cannot be directly compared. To make the results comparable, the Fourier transform was applied to the solution $\mm u(t)$ of
\eref{shooting_evp}. The modal mass \g{\uhone\herm\mm M\uhone} of
the fundamental Fourier component \g{\uhone} equals the square
\g{\amod^2} of the modal amplitude in accordance with \eref{complex_evp}. Likewise, the
harmonics \g{\psihn} of the eigenvector can be obtained. Moreover,
it is useful to a posteriori derive other meaningful quantities from
the frequency domain and the time domain results. In particular, the
maximum value of a specific coordinate or the maximum kinetic energy
were determined. The Jacobian of \eref{shooting_evp} was determined
simultaneously to the actual integration process by integrating the
derivative at each time step. This procedure is well-described \eg
in \zo{seyd1994}.

\subsubsection*{Stability analysis}
The asymptotic stability of the periodic motions was analyzed in
accordance with Floquet's theory. This involved computing the
eigenvalues of the monodromy matrix. It should be noticed that this
matrix is readily available from the computation of the Jacobian of
\eref{shooting_evp}.

\section{Numerical examples\label{sec:numex}}
In order to validate the proposed methodology and to assess its
computational performance, it was applied to various test cases.
Three of these test cases are presented here, namely a van-der-Pol
oscillator, a geometrically nonlinear two-degree-of-freedom system
with viscous damping and a friction-damped rod.

\subsection{Van-der-Pol oscillator\label{sec:vdp}}
First, the van-der-Pol oscillator is considered with the following
equation of motion
\e{\ddot u(t) + \left(\frac{1}{10} u^2(t)-\alpha\right)\dot
u(t)+u(t) = 0\fp}{vdp}
The purpose of this case study is to show the differences
between periodic motion and damped motion concept of nonlinear
modes, and to demonstrate the limitations of the proposed technique
in the presence of strong damping.

\subsubsection*{Modal characteristics}
A common approach to approximate the dynamic behavior of the
van-der-Pol oscillator is the so-called \myquote{harmonic
linearization} technique. To this end, the displacement and the velocity
are approximated in the form \g{u(t)\approx \amod\cos\ommod t} and
\g{\dot u(t)\approx -\ommod\amod\sin\ommod t}, respectively.
Inserting these expressions into \eref{vdp} and truncating the
resulting expressions to their fundamental harmonic component gives
rise to amplitude-dependent equivalent stiffness and damping values,
see \eg \zo{magn1997a}. Just like in the case of a linear
oscillator, approximate undamped eigenfrequency and damping ratio
can thus be determined. For this case, this procedure yields
\e{\ommod(\amod)\approx 1\fk\quad
\dmod(\amod)\approx\frac{\amod^2}{80}-\frac{\alpha}{2}\fp}{vdp_shbm}
It should be noted that throughout this article, $\dmod$ denotes the modal damping ratio and is related to $\deltamp$, as defined in \eref{eqm_periodic}, by $2\dmod\ommod=\deltamp$.\\
It can be ascertained from \eref{complex_evp} that the proposed
Harmonic Balance implementation of the periodic motion concept
degenerates to the harmonic linearization approach for the special
case of a single-degree-of-freedom oscillator if only the
fundamental harmonic component \g{n=1} is considered.
It was verified that the results for both approaches are identical
in the case of \g{\nh=1}, so that only one result will be
shown in the following.\\
Furthermore, the governing equations \erefo{complex_evp} based on the Harmonic
Balance procedure are similar to the ones derived with the
generalized Fourier-Galerkin method in \zo{laxa2009}. The methods
even produce the same results in the special case when \g{\dmod=0},
\ie at limit cycles of the original autonomous system. This
mathematical similarity is remarkable since the authors of
\zo{laxa2009} followed a damped motion concept of the nonlinear
modes, in contrast to the periodic motion concept pursued in the
present study. Their method was also applied to analyze the
van-der-Pol oscillator for comparison.
\figw[tbh]{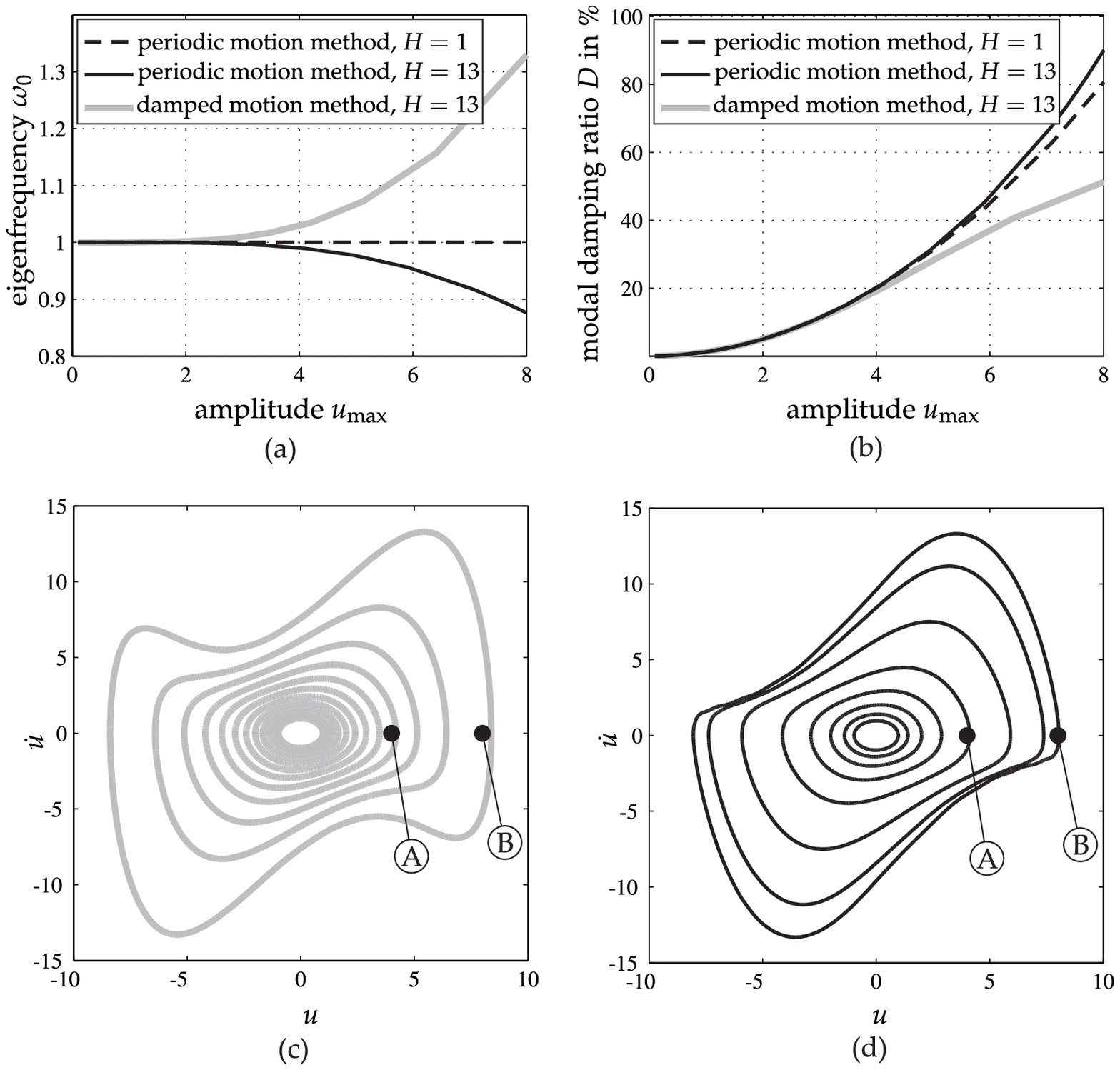}{Modal characteristics of the
van-der-Pol oscillator for $\alpha=0$: (a) eigenfrequency, (b) damping ratio, (c)
phase curves according to damped motion concept with \g{\nh=13},
(d) phase curves according to periodic motion concept with
\g{\nh=13}}
\\
The modal characteristics of the van-der-Pol oscillator for a zero
viscous damping coefficient of \g{\alpha=0} are depicted in
\fref{figure_03}. The results were computed by
solving \eref{complex_evp} for the case of harmonic orders \g{\nh=1}
and \g{\nh=13}. They were also computed by the damped motion method,
\ie the generalized Fourier-Galerkin method which is based on the
damped motion concept \zo{laxa2009}. The amplitude
\g{\umax} is the maximum time-domain displacement (considering all
harmonics). Apparently, the results differ for large
amplitudes, \ie when the nonlinear effects become important. The
frequency-amplitude characteristics exhibit strong differences in
the regime of larger damping, \eg \g{\dmod>20~\%}. The same is true
for the phase curves depicted in \fref{figure_03}c-d.
It should be noticed that these results only indicate that damped and periodic motion concept lead to significantly different results. However, it is not a priori clear which modal characteristics are more \myquote{accurate} in a suitable sense. This aspect is addressed next. In general, the damped motion concept is specifically designed to reflect the strictly
autonomous (decaying or growing) oscillations of the system, while the periodic motion concept
was designed to reflect the (almost) periodic response in the
presence of forced or self-excitation. Two different cases were considered to investigate the ability of the modal characteristics to reflect the dynamic behavior of the nonlinear system: free decay with
\g{\alpha=0}, and limit cycles for varying \g{\alpha>0}.

\subsubsection*{Free decay}
First, the free decay is considered. To approximate
the dynamic behavior using the computed modal characteristics, the
method developed in \zo{krac2014a} was applied. To this end, the
temporal evolution of the amplitude \g{\amod} is governed by the
following differential equation,
\e{\dot\amod\left(t\right) =
-\dmod\left(~\amod\left(t\right)~\right)\ommod\left(~\amod\left(t\right)~
\right)\amod\fp}{freedecay_rom}
\eref{freedecay_rom} was solved using, one after the other, the
modal characteristics \g{\dmod(\amod),\,\ommod(\amod)} obtained by
the periodic motion and the damped motion method. The results are
depicted in \fref{figure_04} and compared to the reference
results obtained from direct time step integration of \eref{vdp}. The
reference envelope (dashed line) is defined as \g{\pm\left(u+\dot
u\right)}.
\figw[tbh]{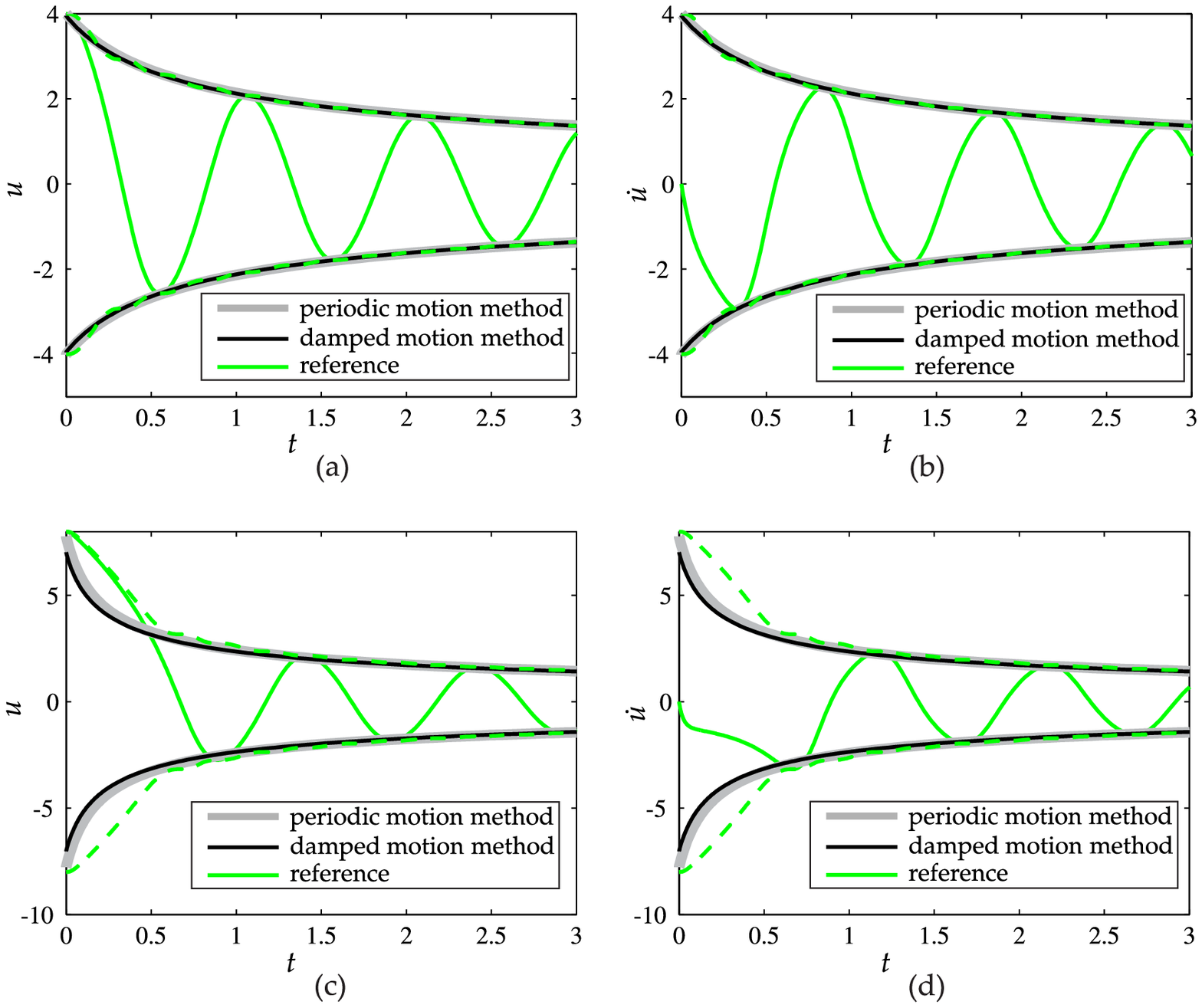}{Time history of the free decay
of the van-der-Pol oscillator: (a) displacement and (b) velocity for
the point A, (c) displacement and (d) velocity for the point B
indicated in \fref{figure_03}c-d}
\\
The most significant
differences are expected in the regime of considerable nonlinear
damping, because here \g{\ommod(\amod)} and \g{\dmod(\amod)} deviate from each other, see \fref{figure_03}. However, in this regime the envelopes are almost indistinguishable. For lower initial energies (point A), the
agreement with the reference is reasonably good. In contrast, the accuracy is
poor for larger initial energies (point B), regardless of the input modal characteristics
\g{\ommod(\amod)} and \g{\dmod(\amod)}. The reason for this deviation is that \eref{freedecay_rom} governs only the approximate slow dynamics nonlinear mode, since an averaging formalism was applied for fast time scale related to the
oscillation with frequency \g{\ommod} \zo{krac2014a}.

\subsubsection*{Limit cycles for varying linear damping}
Next, limit cycles for varying self-excitation intensity, \ie
varying \g{\alpha} were investigated. The steady-state reference results
were obtained by time integration of \eref{vdp}. The nonlinear mode based approximation
was determined by solving the following algebraic equation with
respect to the modal amplitude \g{\amod} \zo{krac2013a}:
\e{0=2\dmod\left(\amod\right)\ommod\left(\amod\right)-\alpha\fp}
{lco_rom}
Herein, \g{\dmod(\amod),\ommod(\amod)} are the modal characteristics
computed for \g{\alpha=0}. It is thus assumed that nonlinear and
linear damping can be \myquote{superimposed} as it is done in
\eref{lco_rom}. \eref{lco_rom} was solved using, one after the other, the modal characteristics obtained from the
respective periodic motion based and the damped motion based method.
\figw[tbh]{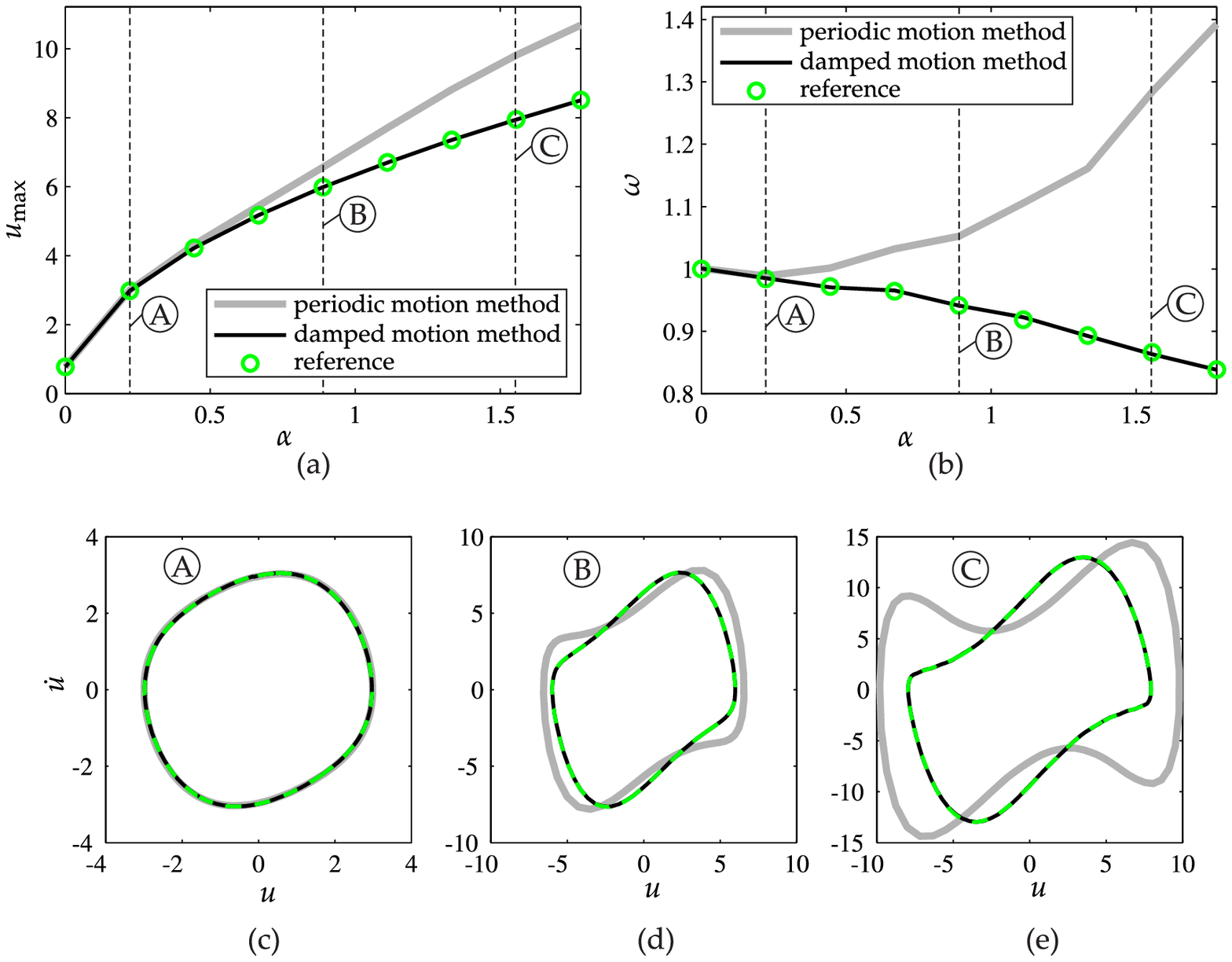}{Limit cycle oscillations of the van-der-Pol
oscillator for varying linear damping $\alpha$: (a) amplitude, (b)
frequency, (c)-(e) phase curves for \g{\alpha = 4/9,\,8/9,\,15/9}
respectively}
\\
The limit cycle amplitude \g{\umax}, frequency \g{\omega} and phase
curve are illustrated in \fref{figure_05} for a varying damping
coefficient \g{\alpha}. The results based on the periodic motion concept agree perfectly with the reference results.
This is expected since the
artificial damping term \g{\deltamp\mm M\mm{\dot u}} introduced for
the modal analysis in \eref{eqm_periodic} has the same effect as the
damping term \g{\alpha\dot u} in \eref{vdp}. In contrast, the results based on the damped motion concept exhibit a poor agreement for \g{\alpha\gg0}. The reason
for this deviation is that damped motions for \g{\alpha=0} were
assumed for the modal analysis rather than limit cycles. In fact,
even the qualitative dependence of the limit cycle frequency on
\g{\alpha} is not correctly predicted.\\\\
In conclusion, the proposed approach performs better than the
conventional one in situations where one is actually interested in
the steady flow of nonconservative systems with a
permanent excitation source. On the other hand, damped motions are -
by design - not very well reproduced by the proposed method.

\subsection{Two-springs system with viscous damping\label{sec:twosprings}}
The system depicted in \fref{figure_06} is now considered. It
consists of a single point mass attached to the ground via two
linear springs. The system can undergo large deflections resulting
in a geometric nonlinearity. This well-known problem was
already investigated \eg in \zo{bell2005,touz2006,blan2013,renson2014}. The
motions of the system are governed by the following set of ordinary
differential equations:
\ea{\nonumber\ddot u_1+2\done\omone\dot u_1
+\omonesq u_1+\quad\quad\quad\quad\\
\quad\quad\frac{\omonesq}{2}\left(3u_1^2+u_2^2\right)+
\omtwosq u_1u_2+\frac{\omonesq+\omtwosq}{2}u_1\left(u_1^2+u_2^2\right)&=&0\\
\nonumber\ddot u_2+2\dtwo\omtwo\dot u_2
+\omtwosq u_2+\quad\quad\quad\quad\\
\quad\quad\frac{\omtwosq}{2}\left(3u_2^2+u_1^2\right)+
\omonesq u_1u_2+\frac{\omonesq+\omtwosq}{2}u_2\left(u_1^2+u_2^2\right)&=&0
\fp}{eqm_twosprings}
If the nonlinear terms are neglected, the system features two modes,
one where the point mass oscillates vertically with the natural
frequency \g{\omone}, and one where it oscillates horizontally with
the natural frequency \g{\omtwo}. For sufficiently small amplitudes,
the motions in these two orthogonal directions can be regarded as
decoupled. As the amplitudes grow, the geometric
nonlinearity causes a coupling between these motions \zo{touz2006}.
As in \zo{touz2006}, linear viscous damping terms are considered. An
interesting feature of the system is that for certain parameter
ranges, damping may change the nonlinear system behavior from a hardening to a softening behavior \zo{touz2006}; \ie, the qualitative dynamic behavior is non-trivial and cannot be explained by the underlying conservative system alone.
This makes the system an ideal test case for the validation of the
proposed methodology. 
\fig[tbh]{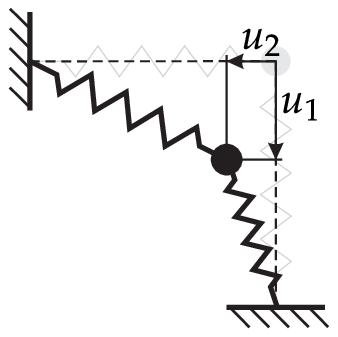}{Model of a geometrically nonlinear
oscillator with two linear springs}

\subsubsection*{Modal characteristics}
The modal characteristics of the first nonlinear mode were computed
using the Harmonic Balance and the Shooting method, \ie by solving
\eref{complex_evp} and \eref{shooting_evp}, respectively. The
results are depicted in \fref{figure_07} for
the parameters listed below the figure. The Harmonic Balance results
are labeled with a \myquote{HB}. The frequency-energy
dependence and the geometry of the invariant manifold generally
resemble previously reported results obtained by normal form theory
\zo{touz2006} or a Galerkin-based approach for computing the
invariant manifold \zo{renson2014}. In particular, the
eigenfrequency first increases slightly with the kinetic energy and
then decreases again. It should be remarked that in absence of damping,
only a hardening effect would be observed. Hence, the damping has an
essential influence on the qualitative vibration behavior. It should be noticed that the modal damping ratio is comparatively small, see \fref{figure_07}b. This seems counter-intuitive owing to the crucial influence of the damping on the dynamic behavior. Apparently, the distribution of damping plays a more important role than its extent.\\
The use of Harmonic Balance also sheds some light into the
underlying dynamics of the system: For larger energies, an important
nonlinear interaction occurs between the first and the second mode.
In the linear case these modes are close to a $2:1$ internal
resonance, \g{2\omone=2\cdot 1.13 \approx 2=\omtwo}. As a consequence,
the second coordinate participates significantly in the second
harmonic component of the motion. Hence, the
second and higher harmonics play an important role in the qualitative evolution of the
modal properties, as it can be seen in
\fref{figure_07}a-b. Moreover, this can be ascertained from the
invariant manifolds depicted in
\fref{figure_07}c-d: During one cycle of oscillation, the coordinate \g{u_2} assumes two maxima and two minima, while the coordinate \g{u_1} only assumes one maximum and one minimum.
\figw[tbh]{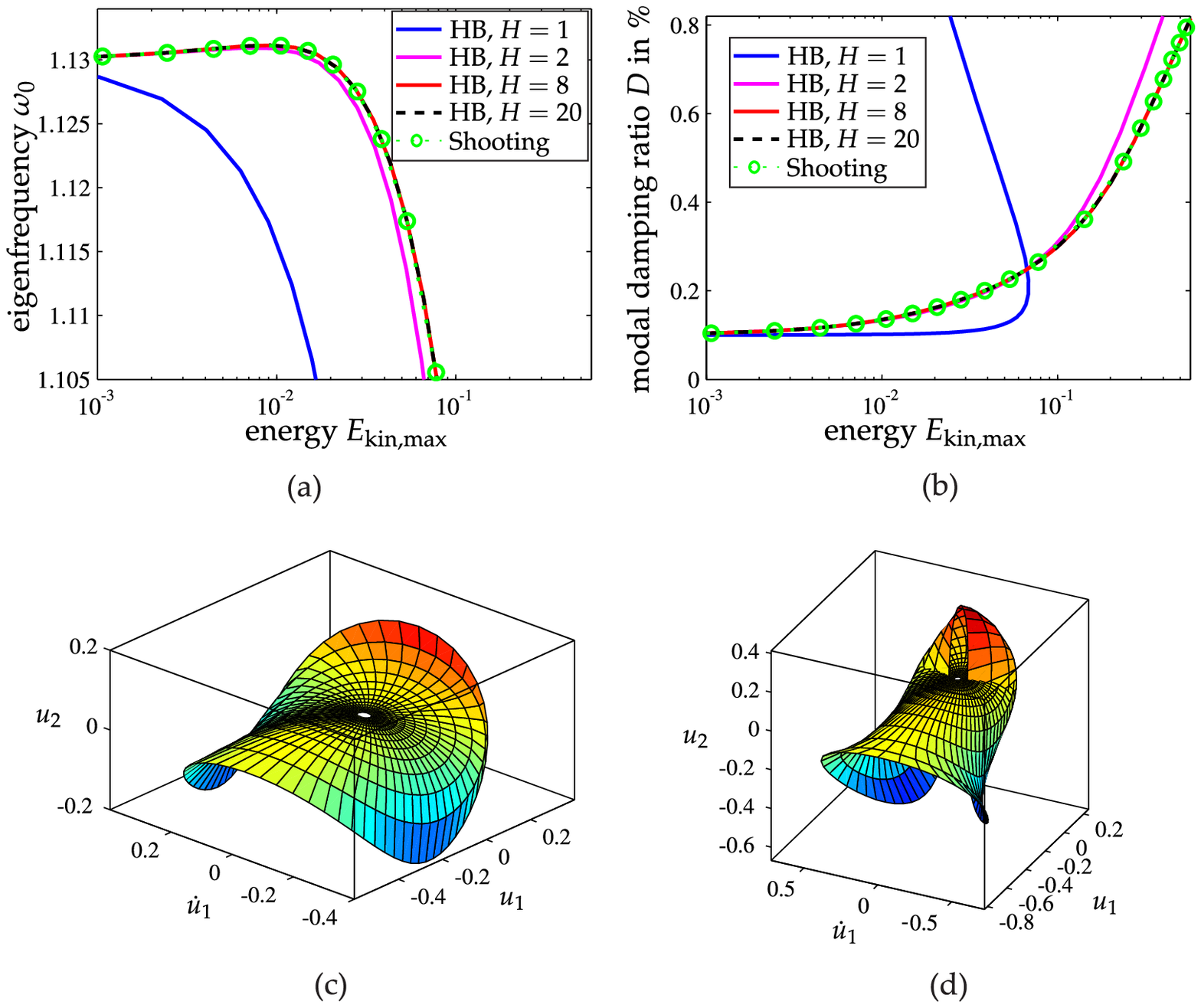}{Modal characteristics of
the first mode of the two-springs system: (a) eigenfrequency, (b)
damping ratio, (c) and (d) invariant manifold prior to and beyond the
fold, respectively; \g{\omone=1.13,\omtwo=2,\done=10^{-3},\dtwo=5\cdot10^{-3}}}
\\
As expected, the results obtained by Harmonic Balance are
indistinguishable from the ones obtained by Shooting for a
sufficiently large number of time steps per period and a
sufficiently large number of harmonics. Several harmonics need to be
retained to achieve asymptotic convergence of the modal
characteristics. If only a single harmonic is considered
(\g{\nh=1}), only a softening behavior is predicted. The folding of
the invariant manifold, as indicated in
\fref{figure_07}d, may lead to parametrization
problems if an invariant manifold based approach for the modal
analysis is pursued. In contrast, the folding does not lead to any
special attention when using the proposed procedure.

\subsubsection*{Forced response}
The steady-state forced response of the two-springs system was also investigated. A harmonic
excitation was applied to the point mass in both coordinate
directions, with the same magnitude and equal phase. Hence, additional forcing terms have to be considered on the right hand side of \eref{eqm_twosprings}.
The Shooting method was applied to the resulting equations. The frequency
range around \g{\omone} was considered. A predictor-corrector path
continuation scheme was applied to overcome the turning points.\\
The frequency response curves of coordinate \g{u_1} are depicted in
\fref{figure_08}a-b for different excitation levels. The maxima
of the frequency response curves match well the backbone curve,
which represents the frequency-amplitude characteristic of the
nonlinear mode. This is true for all depicted excitation levels, in
spite of the fact that the resonant response looses its
asymptotic stability for large excitation levels, as indicated in
\fref{figure_08}b, and in spite of the folding of the
invariant manifold at these energy levels, \cf
\fref{figure_07}d. In the presence of damping,
the frequency at the maximum forced response does of course not
exactly coincide with the undamped eigenfrequency. Note that this is
also true in the linear case. However, as it can be ascertained from
\fref{figure_07}b, the modal damping ratio is
comparatively small, so that this difference cannot be seen in
\fref{figure_08}a-b. From the high accuracy of the
frequency-amplitude characteristic, it can be concluded that the
introduction of the artificial damping term in \eref{eqm_periodic}
does not diminish the integrity of the proposed approach for this
test case.
\figw[tbh]{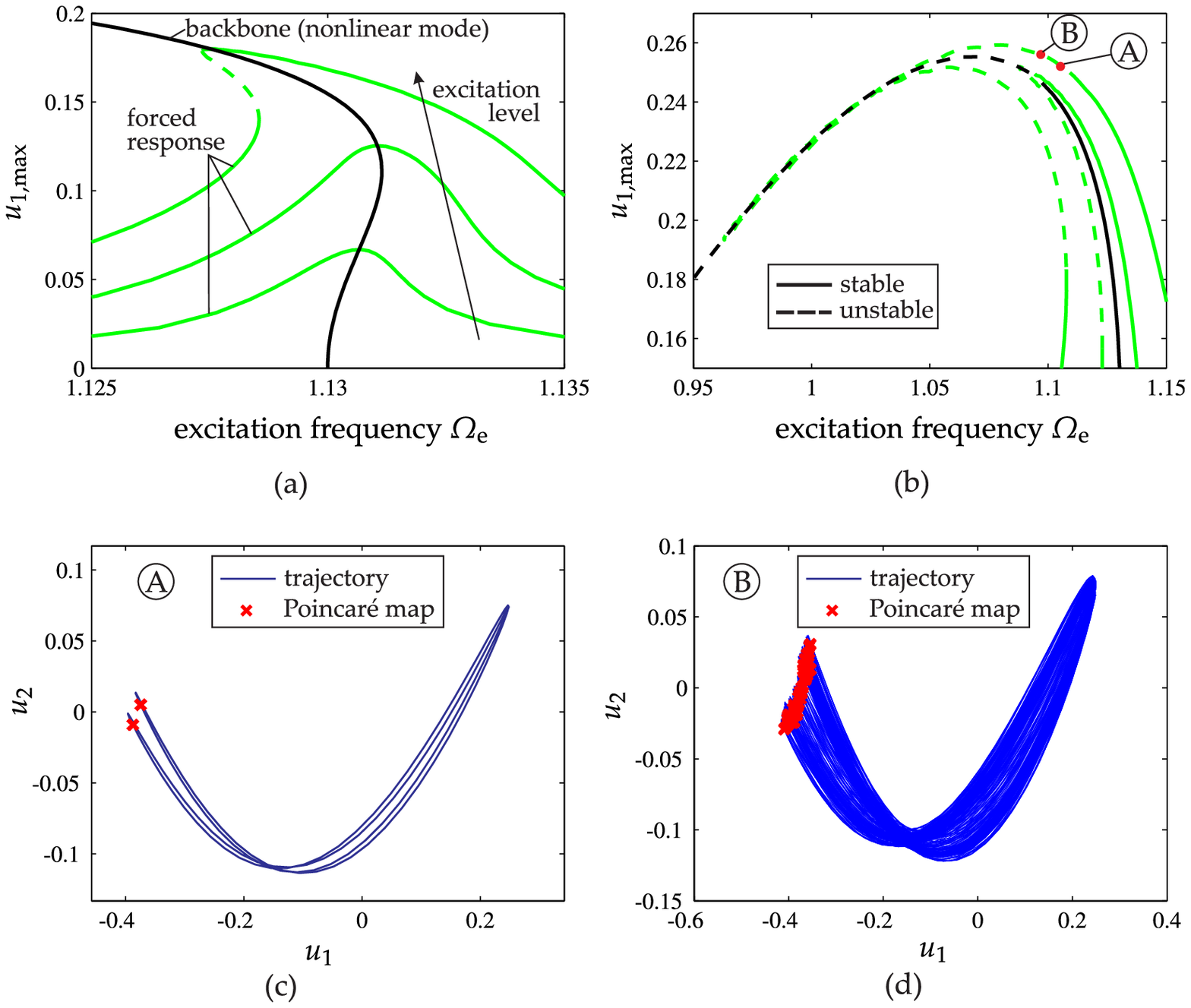}{Forced response of the two-springs
system: (a)-(b) amplitude-frequency curves and backbone for lower
and higher excitation levels, (c)-(d) stable phase projections in
the displacement space for two cases where the one-periodic response
is unstable}
\\
The asymptotic stability of the periodic motions was also
investigated by means of Floquet's theory. To this end, the
eigenvalues of the associated monodromy matrix were analyzed. It
should be noticed that this is a comparatively inexpensive
post-processing step in conjunction with the Shooting method. As
expected, the overhanging part of the frequency response curves
between the two turning points is unstable. At the bifurcation points, a Floquet eigenvalue exits the unit circle in the complex
plane at $+1$ on the real axis. This indicates a saddle node bifurcation. Starting from a point on the unstable branch, a transient motion would be initiated until a steady-state vibration on one of the stable branches is reached.
Since the maximum of the
frequency response curve is not on the overhanging branch, this type
of instability is not present in the backbone curve.\\
Beyond a certain excitation level, a part of the frequency response around the resonance becomes unstable as well. In this case, a Floquet eigenvalue exits the unit circle in the complex
plane at $-1$ on the real axis at the bifurcation points. This indicates a period doubling
bifurcation. To the best of the author's knowledge, this type of
bifurcation was not reported so far for the considered system
\zo{touz2006,renson2014}. To confirm this behavior, the points A and B
in \fref{figure_08}b are considered, which are on the unstable
section of one of the frequency response response curves. Starting
from the corresponding initial state \g{\mm u_0,\mm v_0}, the
transient dynamic behavior was determined by means of direct
time-step integration. After a certain number of transient periods,
the motion was found to be steady. The motion starting from point A
reaches a periodic orbit with frequency \g{2\Omega_{\mathrm e}}, see
\fref{figure_08}c. This is in accordance with the results of
the stability analysis. In contrast, the motion starting from point
B does not reach a periodic orbit but approaches a non-periodic, possibly quasi-periodic
attractor, see \fref{figure_08}d.
Apparently, bifurcations other than the considered flip bifurcation take place
in this regime.\\
Remarkably, the point ($\Omega_{\mathrm e}\approx 1.1$,
$u_{1,\max}\approx 0.24$) and the type of stability loss of the forced
response are in good agreement with the stability analysis of the
nonlinear mode. Apparently, the nonlinear modes defined in \eref{eqm_periodic} capture not only the frequency-energy dependence but also the asymptotic stability of \g{1:1} external resonances.

\subsection{Friction-damped rod\label{sec:fricrod}}
Friction damping is a well-known means of accomplishing vibration reduction \zo{ferr1995,popp2003a}. The damping effect
is achieved by the dissipative character of dry friction occurring in mechanical joints. Friction damping is particularly
suited for the passive vibration reduction of lightly damped structures. Various applications
can be found in the field of aerospace structures, combustion engines or turbomachinery
blades. For the design of a friction-damped system, different performance measures are of particular interest:
In externally forced scenarios, the resonance frequency shift is crucial for the detection
and avoidance of possible resonances \zo{elli2008a}. In
self-excited scenarios, the effective damping ratio is often assessed
\zo{sinh1983,whit1996b}. The proposed method is well-suited to determine these performance measures directly and thus represents an apt alternative to the crude forced response simulations widely-used in this context.\\
A simple model of a slender friction-damped rod illustrated in \fref{figure_09} is considered.
The rod of length $\ell$ is clamped at one end and it is connected
to an elastic Coulomb dry friction element at its other end. Only
small longitudinal vibrations of the rod were assumed, and the
material was considered as homogeneous and linear elastic, with a Young's modulus $E$ and a density $\rho$. The equations of motion of the continuous system can be summarized as follows:
\ea{\text{local equilibrium} &\rho\frac{\partial^2 u}{\partial t^2}(x,t) - E\frac{\partial^2 u}{\partial x^2}(x,t) = 0\fk\,\,0< x < \ell\fk\\
\text{fixed boundary} & u(0,t) = 0\fk\\
\text{friction contact} & EA\frac{\partial u}{\partial x}(\ell,t) = -\fr\left[u(\ell,t)\right]\fk\\
\text{elastic Coulomb law} & \dd\fr =\begin{cases}\kt\dd u & \text{if}\,\,\left|\fr+\kt\dd u\right|<\mu N\\ 0 & \text{otherwise}\end{cases}\fp
}{rod_eqm}
Note that the elastic Coulomb law for the friction force $\fr$ is written in differential form, where $\kt$ denotes the contact stiffness and $\mu N$ is the limit friction force.\\
The structural dynamic behavior of the rod was approximated using a
finite element procedure. To this end, $20$ rod elements of equal
length \g{\ell_{\mathrm e}=\ell/20} were defined, each having the
following consistent element mass and stiffness matrices \zo{cook2002}:
\e{\mm M_{\mathrm e} = \frac{\rho A\ell_{\mathrm e}}{6}\matrix{cc}{2
& 1\\ 1 & 2}\fk\quad \mm K_{\mathrm e} = \frac{EA}{\ell_{\mathrm
e}}\matrix{cc}{1 & -1\\ -1 & 1}\fp}{rod_elementmatrices}
Herein, \g{A} is the cross section area.
\fig[tbh]{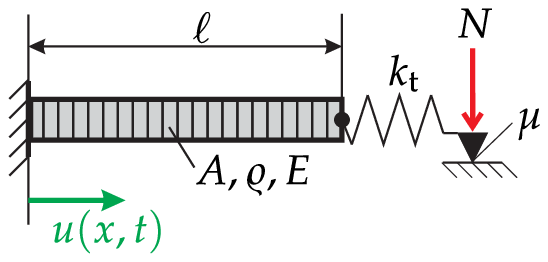}{Model definition of the friction-damped rod
undergoing longitudinal vibrations}

\subsubsection*{Modal characteristics}
The first nonlinear mode was analyzed
using the Harmonic Balance and the Shooting method, \ie by solving
\eref{complex_evp} and \eref{shooting_evp}, respectively. The energy
dependence of the modal frequency, damping ratio and the invariant
manifold is depicted in \fref{figure_10}.
Throughout the figures in this subsection, some quantities are
normalized which is indicated with an asterisk ($*$). Frequencies
were normalized by the eigenfrequency in the linear case with
sticking contact.
\figw[tbh]{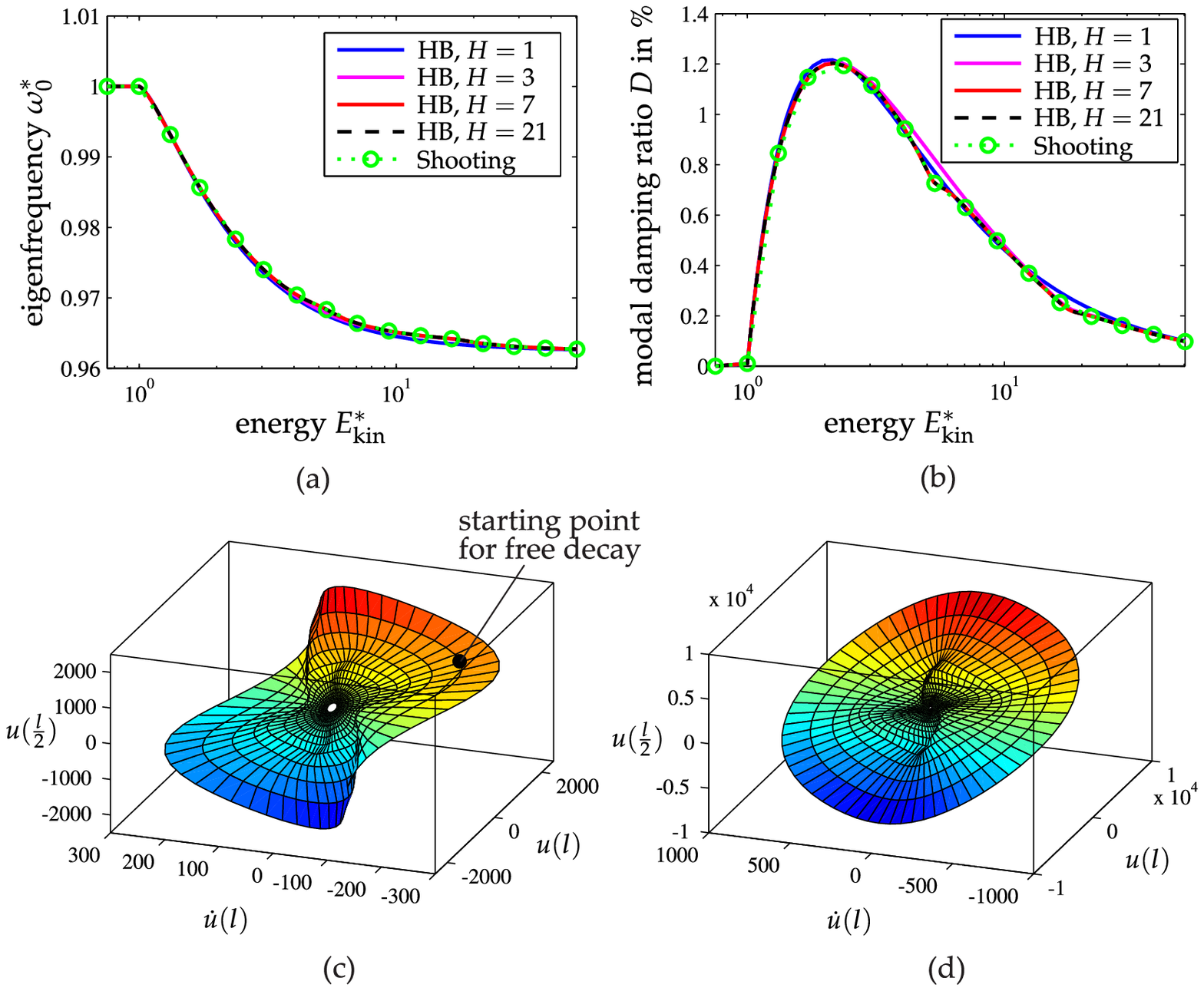}{Modal characteristics of
the first longitudinal mode of the friction-damped rod: (a)
eigenfrequency, (b) damping ratio, (c)-(d) invariant manifold for
different energy ranges; \g{\ell=20,A=1,\rho=1,E=1,\kt=0.005,\mu N=1}}
\\
For low kinetic energies, the friction element is only sticking,
which results in a constant eigenfrequency and zero damping. Beyond
this linear regime, the friction contact undergoes slipping in
addition to sticking. The kinetic energy \g{E_{\mathrm{kin}}} was
normalized by its value at the end of the linear energy regime, \ie
when slipping occurs for the first time instead of only sticking.
The slipping phases cause a softening and a damping effect. For
large energies, sliding friction dominates, and the modal frequency
approaches the eigenfrequency of the linear rod without friction
contact. The dissipated energy grows linearly with the displacement
amplitude. For comparison, it would grow quadratically for a linear
viscous damper, for which \g{\dmod} would be constant. Hence, the
effective damping ratio \g{\dmod} of the friction-damped system
tends towards zero. The modal characteristics qualitatively resemble
the results previously reported by other researchers, see \eg
\zo{whit1996,zasp2007,laxa2009}.\\
In \fref{figure_10}c-d, the two-dimensional invariant manifold is
depicted for different energy levels in a three-dimensional subspace
of the \g{40}-dimensional phase space. The closed curves correspond
to the computed periodic motions. Apparently, the orbits deviate
from ellipses, which indicates higher harmonic content of the
periodic motions. Moreover, the orbits do not lie in a plane
anymore, which indicates that the mode shape also depends on the
kinetic energy. This behavior is certainly influenced by the non-smooth character of the dry friction forces.\\
The results obtained from the Shooting and the Harmonic Balance
method are in good quantitative agreement. It should be noticed that
frequency-energy and damping-energy characteristic do not
significantly depend on the harmonic order \g{\nh}, even though the
multi-harmonic character of the periodic motions can be easily
ascertained from the invariant manifold.

\subsubsection*{Free decay}
The free decay of the first nonlinear mode was investigated next.
The starting point is indicated in
\fref{figure_10}c. The temporal evolution of the
modal amplitude was approximated by numerical integration of
\eref{freedecay_rom}. For reference, the time history of the
coordinates was determined by numerical integration of the full set
of equations of motion \erefo{eqm_autonomous}.
\fig[tbh]{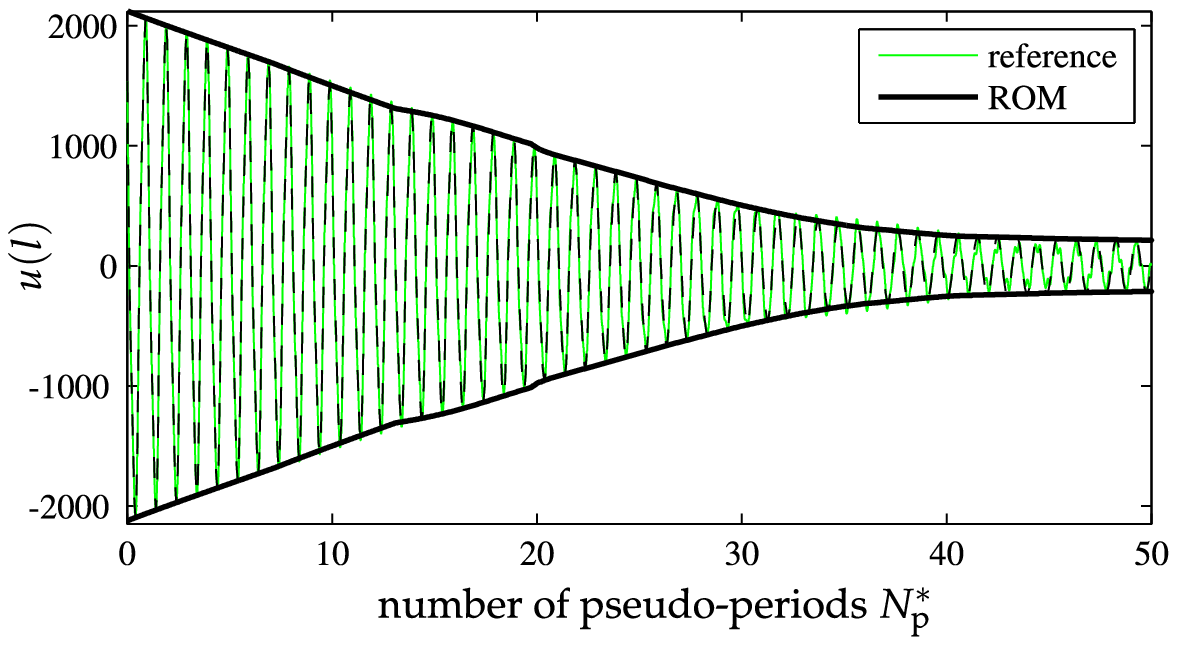}{Time history of the free decay from the
starting point indicated in \fref{figure_10}c}
\\
The results for the evolution of the rod's tip displacement
\g{u(\ell)} are shown in \fref{figure_11}. In addition to the envelope, the synthesized temporal evolution of \g{u(\ell)} is depicted for the nonlinear mode based approximation. Details about this approximate synthesis can be found in \zo{krac2014a}. The number
\g{\npp} of pseudo-periods was introduced as time variable: It
equals \g{\npp=f_0t}, where \g{f_0} is the eigenfrequency of the
considered first mode in the linear case. The results are in
reasonably good agreement. Some deviations can be encountered in
particular starting from \g{\npp>35}. This is the time range
slightly before and after the friction element is in permanently
sticking configuration and the amplitude remains constant. It is
conjectured that the non-smooth state transitions between stick and
slip are particularly important in this regime and excite other
frequencies. This effect cannot be well-resolved with the single nonlinear mode approximation.

\subsubsection*{Forced response}
Finally, the steady-state forced response to harmonic excitation was
investigated. A harmonic forcing was applied to the tip
of the rod. The forced response was computed by means of the
Shooting method applied to the full set of equations of motion in
the forced setting.
\fig[tbh]{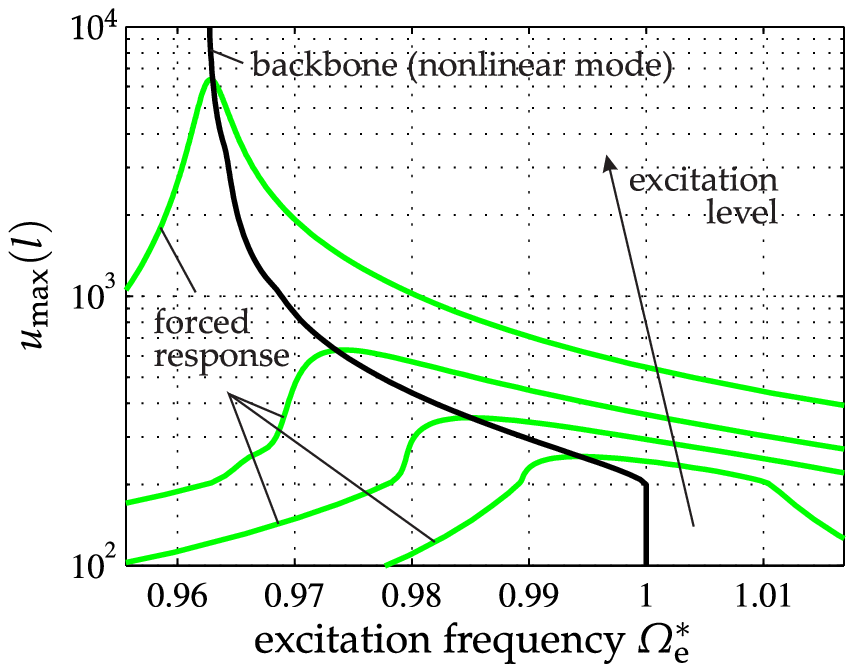}{Amplitude-frequency curves of the
harmonically forced friction-damped rod and backbone obtained from
nonlinear modal analysis}
\\
In \fref{figure_12}, the frequency response curves are depicted
for varying excitation level around the first eigenfrequency. The amplitude $u_{\max}(l)$ is defined
as the maximum time domain displacement of the rod's tip coordinate
$u(\ell)$. The forced response results confirm the softening effect.
Moreover, they indicate the decrease of the damping effect, which
results in a narrower resonance peak for larger energy levels. The
maxima of the frequency response curves also agree quantitatively
with the frequency-amplitude characteristic of the nonlinear mode,
\ie the backbone curve.

\section{Evaluation of the computational performance}
For the two-springs system and the friction-damped rod presented in
\sref{twosprings} and \sref{fricrod}, respectively, the
computational effort was investigated. All numerical procedures were
implemented and carried out in Matlab R2012a. The simulations were carried out
on an Intel Core 2 Quad Q9550 $2.83~\mathrm{GHz}$ CPU with
$3.62~\mathrm{GB}$ addressable memory. In
\tref{twosprings_compeffort} and \tref{fricrod_compeffort}, the
computational effort for the nonlinear modal analysis is listed.
\tab[tbh]{lcc|cc}{%
analysis & $\nh$ & $\ntd$ & $\niter$ & computation time in $\mathrm s$\\
\hline
\hbm     & $1$   & $16$   & $44$     & $0.22$\\
\hbm     & $2$   & $16$   & $48$     & $0.25$\\
\hbm     & $8$   & $32$   & $47$     & $0.35$\\
\hbm     & $20$  & $2048$ & $47$     & $0.90$\\
Shooting & -     & $128$  & $102$    & $4.2$\\
Shooting & -     & $256$  & $107$    & $8.4$\\
Shooting & -     & $1024$ & $104$    & $31.3$ }{Computational effort
for the nonlinear modal analysis ($24$
data points) of the two-springs system with $2$ generalized coordinates}{twosprings_compeffort}
\tab[tbh]{lcc|cc}{%
analysis & $\nh$ & $\ntd$ & $\niter$ & computation time in $\mathrm s$\\
\hline
\hbmwocond & $1$   & $16$   & $152$     & $1.3$\\
\hbmwocond & $3$   & $16$   & $191$     & $2.7$\\
\hbmwocond & $7$   & $32$   & $205$     & $11.2$\\
\hbmwocond & $21$  & $2048$ & $206$     & $99.4$\\
Shooting   & -     & $64$   & $280$     & $9.8$\\
Shooting   & -     & $128$  & $302$     & $11.4$\\
Shooting   & -     & $256$  & $282$     & $20.9$\\
Shooting   & -     & $1024$ & $258$     & $68.3$ }{Computational
effort for the nonlinear modal analysis ($80$ data points) of the friction-damped rod with $20$ generalized coordinates}{fricrod_compeffort}
\\
In the case of the Harmonic Balance (HB) method, the computations
were carried out for different harmonic orders \g{\nh}, see \eg
\fref{figure_07}. In general, the number
\g{\ntd} of sampling points in the alternating-frequency-time scheme
should be selected as large as necessary to ensure sufficient
accuracy and as small as possible to avoid spurious computational
effort. The optimal number \g{\ntd} is typically not a priori known
and may depend on the harmonic order \g{\nh}. For the smaller
harmonic orders \g{1\leq\nh\leq 8}, \g{\ntd} was selected as the
smallest power of two beyond which frequency and damping
characteristics did not change significantly anymore. For the
largest harmonic order \g{\nh=20} and \g{\nh=21}, respectively,
\g{\ntd} was set to an unnecessarily large value, just to show its
influence on the computation time.\\
In the case of the Shooting method, the number \g{\ntd} of time
steps in the implicit average acceleration Newmark scheme was
varied. Only powers of two were specified. The smallest of
the listed numbers was considered just large enough to avoid poor
qualitative agreement with the converged results. In order to ensure
that the results remain comparable, the same number of data points,
\ie modal amplitudes was considered in the modal analyses. $24$ data
points were specified in the case of the two-springs system and $80$
data points in the case of the friction-damped rod.\\
For both test cases, sufficiently accurate results can be produced
within several seconds. For almost all analyses, the computation
time remained below one minute. In other studies \zo{renson2014,renson2014b} considered similar test cases, but reported computation
times of several minutes using a Galerkin-based invariant
manifold approach. It is therefore believed that the proposed
periodic motion based method is computationally less expensive
than invariant manifold based numerical methods. This conclusions
seems plausible owing to the typically smaller problem dimension
involved in each iteration of the respective approaches, \cf the
comment in the introduction.
\\
From a computational point of view, the Harmonic Balance method was
considerably less expensive than the Shooting method for the
calculations carried out in this study. The benefit of Harmonic
Balance appears to be due to both the smaller number of required
iterations and the lower computational cost per iteration.

\section{Conclusions\label{sec:conclusions}}
A novel definition of the nonlinear modes of nonconservative systems
was proposed. They were defined as self-excited limit cycles induced
by (typically negative) mass-proportional damping. While the conventional damped motion concept aims at reflecting the strictly autonomous dynamics, the proposed periodic motion concept is designed to capture the periodic dynamics induced by a resonant excitation source. In a wide range, the numerical examples demonstrated that the extracted modal properties accurately describe the vibration behavior in the vicinity of forced resonances and self-excited limit cycles. This applies to the modal frequency, damping ratio, vibrational deflection shape as well as the asymptotic stability. The limitations of the concept are due to the intrusiveness of the artificial damping term. The concept is fully consistent with the conservative nonlinear or the nonconservative linear case. In the nonconservative nonlinear case, however, the artificial damping term may induce additional modal coupling. Hence, the proposed concept is restricted to the dynamic regimes of isolated nonlinear modes and/or regimes with reasonably low modal damping ratios. Moreover, the definition is limited to periodic motions, and therefore non-periodic regimes which may also occur near resonance, are generally not captured.\\
An important benefit of the proposed concept is that standard methods for analyzing periodic motions can be applied.
Two computational procedures for the modal analysis were described
and investigated in this study. One of them is based on the Harmonic
Balance method and one on the Shooting method. It is noteworthy that the Harmonic Balance based procedure degenerates to the well-known harmonic linearization technique if the model comprises only a single degree of freedom. Both procedures are
applicable to generic nonlinearities, including those involving
non-smooth forces. Their computational performance is comparatively good, even if the model features a larger number of generalized coordinates.\\
The overall approach is regarded as an effective and computationally efficient tool for modal parameter extraction of nonconservative nonlinear systems in the aforementioned dynamic regimes. Since the energy-dependent modal properties accurately reflect the resonant vibration behavior, they represent relevant measures for the structural dynamic design of nonlinear structures. For instance, the proposed approach gives directly rise to a measure of the damping effectiveness.
This is obviously a crucial feature for the design assessment of vibration reduction mechanisms such as friction damping. As it was demonstrated in previous works \zo{krac2013a,krac2014a}, the modal characteristics can also be utilized for model reduction. This facilitates quantitatively accurate vibration predictions
for a wide range of operating conditions.

\end{document}